# Fractals and Fractal Dimension of Systems of Blood Vessels:

## An Analogy between Artery Trees, River Networks, and Urban Hierarchies

Yanguang Chen

(Department of Geography, College of Urban and Environmental Sciences, Peking University, Beijing 100871, P.R. China. E-mail: chenyg@pku.edu.cn)

**Abstract**: An analogy between the fractal nature of networks of arteries and that of systems of rivers has been drawn in the previous works. However, the deep structure of the hierarchy of blood vessels has not yet been revealed. This paper is devoted to researching the fractals, allometric scaling, and hierarchy of blood vessels. By analogy with Horton-Strahler's laws of river composition, three exponential laws have been put forward. These exponential laws can be reconstructed and transformed into three linear scaling laws, which can be named *composition laws of blood vessels network*. From these linear scaling laws it follows a set of power laws, including the three-parameter Zipf's law on the rank-size distribution of blood vessel length and the allometric scaling law on the length-diameter relationship of blood vessels in different orders. The models are applied to the observed data on human beings and animals early given by other researchers, and an interesting finding is that human bodies more conform to natural rules than dog's bodies. An analogy between the hierarchy of blood vessels, river networks, and urban systems are further drawn, and interdisciplinary studies of hierarchies will probably provide new revealing examples for the science of complexity.

# 1 Introduction

Coming from and evolving in geographical environments, creatures must follow the natural laws of geographical environments. At least, creatures and environments may follow the same natural laws. Thus, there should be correlation and similarity between organisms and geographical environments. In fact, ancient Chinese sages such as Laozi recognized the relationships between human and the earth. After the Second World War, many studies on creatures and its environments emerged from time to time. For example, a similarity between the shapes of the abundance curves for the elements present in human tissues and those in average crustal rocks is discovered, which indicates the corresponding relationships between the chemical elements present in man and those in the natural environment (Hamilton *et al*, 1973). An interesting studies on hierarchical systems with cascade structure such cities, rivers, Alpine glaciers, bovine livers and trees was once made by Woldenberg (1968), and this research suggests the similarities between animal's organs and river networks or even city systems. Due to fractal theory (Mandelbrot, 1982), more analogies were drawn between creatures and geographical phenomena (Batty and Longley, 1994; MacDonald, 1983; Nelson and Manchester, 1988).

There is an analogy between the structure of hierarchies of blood vessels of mammals and that of network of rivers and streams. Many studies showed that river and stream networks bear fractal properties (e.g. Chen and Li, 2003; LaBarbera and Rosso, 1989; Rodriguez-Iturbe and Rinaldo, 2001; Turcotte, 1997). The laws of river composition proposed by Horton (1945) and developed by Strahler (1952) and Schumm (1956) suggest self-similar structure, and the law of Hack (1957) suggests allometric scaling relation (Chen and Liu, 2001). By analogy with river networks, the fractal nature of blood vessel hierarchies has been researched, and Horton-Strahler's laws have been employed to describe the structure of human and mammals' arterial trees (Finet *et al*, 2007; Jiang and He, 1989; Jiang and He, 1990; Liu *et al*, 1992; Wang *et al*, 2001; Ying *et al*, 1993). These studies verified the correlation and similarities between creatures and geographical environments. However, many basic problems remain to be further explored by theoretical derivations and empirical analyses. On the one hand, the deep structure of self-similar hierarchies of blood vessels is not yet modeled using fractal geometric geometry; on the other hand, the mathematical essence of blood vessel trees is not yet brought to light by fractal dimension.

Recent years, progress has been made in studies on fractal hierarchy of cities (Batty and Longley, 2010; Chen, 2008; Chen, 2011; Chen, 2012; Chen, 2014; Chen and Zhou, 2003; Jiang and Yao, 2010; Pumain, 2006). There exists significant analogy between the hierarchies of cities and towns and the networks of rivers and streams (Chen, 2009; Chen and Zhou, 2006; Krugman, 1996; Woldenberg and Berry, 1967). This suggests that some new findings in urban studies can be applied to modeling blood vessel hierarchies, and in turn the advance made in the studies on blood vessels will be returned to the studies on rivers and cities. This paper is devoted to researching fractals and fractal dimension of systems of blood vessels on the analogy of river networks and urban hierarchies. In Section 2, the models of hierarchies of blood vessels are reconstructed by analogy with the laws of river composition. From the linear scaling laws (exponential laws) it follows a series of nonlinear scaling laws (power laws). In Section 3, a case study on human and dogs' arterial trees in a previous work is reinterpreted using the ideas from fractals. A formula for adjusted fractal dimension of hierarchies of blood vessels is presented. In Section 4, several related questions are discussed, and the scaling laws of blood vessels are developed and generalized to other fields such as geomorphology and urban geography. Finally, the discussion is concluded by highlighting the main points of this study.

## 2 Models: Analogy between blood vessels and rivers

### 2.1 The laws of blood vessel network composition

The basic mathematical laws of blood vessels can be given by exponential functions. Previous studies show that the bifurcation number, average length, and average diameter of creatural arteries take on exponential growth or exponential decay (Jiang and He, 1989; Jiang and He, 1990; Li *et al*, 1992). Let's consider a hierarchy of blood vessels consisting of $M$ levels. Based on the bottom-up method of hierarchy recognition, the average diameter change of blood vessels of different levels can be expressed as follows

$$d_m = d_0 e^{\rho m}, \tag{1}$$

where $m$ denotes the order of the blood vessel level ($m=1, 2,…, M$), $d_m$ refers to the average diameter of the blood vessels of order $m$, and $d_0$ and $\rho$ are two parameters: the former is a proportionality parameter, and the latter is a rate parameter. The average length change of blood vessels can be expressed as blow

$$h_m = h_0 e^{\sigma m}, \tag{2}$$

in which $h_m$ denotes the average length of the blood vessels of order $m$, and $h_0$, $\sigma$ are proportionality and rate parameters, respectively. The bifurcation number of blood vessels can be expressed in the following form

$$N_m = N_0 e^{-\omega m}, \tag{3}$$

where $N_m$ refers to the bifurcation number of the blood vessels of order $m$, and $N_0$ and $\omega$ are two parameters indicative of proportionality and rate. The reciprocal of the rate parameter suggests a scale parameter.

The above exponential laws are in fact three linear scaling laws of hierarchical structure (Lee, 1999; Williams, 1997). Equations (1), (2), and (3) represents the basic mathematical laws of blood vessel networks. Equation (1) bears an analogy with the exponential model of the spiral structure of trumpet shells. Thompson (1917) found that the radius of a trumpet shell depends on the corresponding angle, which suggests a linear scaling law (Lee, 1999). By analogy with the trumpet model, equation (1) can be rewritten as below

$$R_m = R_0 e^{\rho m}, \tag{4}$$

in which $R_m = d_m/2$ denotes a radius, and $R_0 = d_0/2$ refers to a proportionality coefficient. Taking the logarithm of equation (4) yields a linear relation such as $\log R_m = \log R_0 + \rho m$, which indicates a linear scaling. Equations (2) and (3) can be understood in a similar way.

There is an analogy between blood vessels and rivers, and thus there are an analogy between the mathematical laws of blood vessel composition and river composition (Jiang and He, 1989; Jiang and He, 1990; Liu et al, 1992). The law of river composition was proposed by Horton (1945) and developed by Strahler (1952) and Schumm (1956). On the analog of the river composition laws, we can reconstruct the mathematical models of the blood vessels and then derive the nonlinear scaling laws of blood vessel networks. Let $N_1 = N_0 e^{-\omega}$. Equation (3) can be re-expressed as

$$N_m = N_1 e^{-\omega(m-1)} = N_1 r_b^{1-m}, \tag{5}$$

where $r_b = e^{\omega}$, namely, $\omega = \ln r_b$, in which $r_b = N_m/N_{m+1}$ refers to the bifurcation ratio of blood vessels at different levels. Apparently we have, $N_M = N_1 r_b^{1-M}$, and thus $N_1 = N_M r_b^{M-1}$. Inserting this formula into equation (5) yields

$$N_m = N_M r_b^{M-m}, \tag{6}$$

which is identical in form to the first law of Horton's river composition. Here $N_1$ refers to the bifurcation number of blood vessels of order 1 (bottom class), and $N_M = 1$ to the bifurcation number of blood vessels in the $M$th class (top class). Equation (2) can be re-expressed as

$$h_m = h_1 e^{\sigma(m-1)} = h_1 r_l^{m-1}, \tag{7}$$

where $h_1 = h_0 e^{\sigma}$, $r_l = e^{\sigma}$, and accordingly, $\sigma = \ln r_l$. From equation (7) it follows $r_l = h_{m+1}/h_m$, which suggests a mirror symmetry of the hierarchy of blood vessels. Because of the symmetry of exponential laws, if the bottom-up order (from the smallest to the largest) is replaced by the top-down order (from the largest to the smallest), the basic form of the mathematical expression will not change, but the sign of the exponent will change: the positive sign will change to the negative sign. For example, equation (7) can be rewritten as $h_m = h_1 r_l^{1-m}$, in which $m$ denote the top-down order instead of the bottom-up order, and therefore $r_l = h_m/h_{m+1}$.

The branch length law can be replaced by the cumulative length law of hierarchies. Based on equation (7), the cumulative length of blood vessels of different orders can be expressed as

$$L_M = h_1 \sum_{i=1}^{M} r_l^{1-i} = \frac{h_1 r_l}{r_l - 1}, \quad L_{M-1} = h_2 \sum_{i=2}^{M} r_l^{2-i} = \frac{h_2 r_l}{r_l - 1}. \tag{8}$$

As a result, we have a special length ratio $r_l = L_M/L_{M-1} = h_1/h_2$, in which $L_M$ refers to the cumulative length from the bottom level to the top level, and $L_{M-1}$ to the cumulative length from the bottom level to the second highest level. By recursion, we have a general length ratio such as $r_l = L_{m+1}/L_m = h_{m+1}/h_m$. Based on the length ratio, the length law of blood vessel can be expressed as

$$L_m = L_1 r_l^{m-1}, \tag{9}$$

which is identical in form to the second law of Horton's river model. Here $L_m$ denotes the average length of blood vessels from the bottom level to the $m$th level, and $L_1$ indicates the average length of blood vessels at the first level (bottom level). Note that the order number in equation (9) is still from the bottom up. Finally, equation (4) can be rewritten in the following form

$$R_m = R_1 e^{\rho(m-1)} = R_1 r_z^{m-1}, \tag{10}$$

in which $R_m = R_1 e^{\rho}$, $r_z = e^{\rho}$, and accordingly, we have $\rho = \ln r_z$, where $r_z = R_{m+1}/R_m$ represents the caliber ratio. Equation (10) is theoretically equivalent to equation (9), and it is not corresponding to the third law of Horton's river model. This problem will be discussed later. Thus, we have fulfilled the first task of reconstructing the models of blood vessel composition. The result is the three laws:

the first is the bifurcation law expressed by equation (5) or equation (6), the second is the length law expressed by equation (7) or equation (9), and the third is the aperture law expressed by equation (10).

## 2.2 The fractals and allometry of blood vessel networks

In order to understand the fractal nature of blood vessel network, it is necessary to derive the rank-size scaling law and allometric scaling law of the hierarchies of human and mammalian blood vessels. First, change the direction of the serial numbers of the levels in a hierarchy: turn the bottom-up order into the top-down order. That is, 1 is replaced by $M$, 2 is replaced by $M$-1, …, and $M$ is replace by 1. Because of the translational symmetry of exponential distribution, reordering the levels of a hierarchy will not change the basic mathematical form of the linear scaling relation based on exponential functions. Thus equations (5), (9), and (10) will be substituted with the following functions

$$N_m = N_1 r_b^{m-1}, \tag{11}$$

$$L_m = L_1 r_l^{1-m}, \tag{12}$$

$$R_m = R_1 r_z^{1-m}, \tag{13}$$

where $N_m$ denotes the number of blood vessels of the top-down order $m$, $N_1$ is the blood vessel number of the first order (generally speaking, $N_1=1$), $L_m$ refers to the average cumulative length of blood vessels of the top-down order $m$, $L_1$ is the overall length of the top-level blood vessel, $R_m$ indicates the average radius of blood vessels of the top-down order $m$, $R_1$ is the average radius of the primary blood vessel. Thus, we have fulfilled the second task of reconstructing the models of blood vessel composition.

A three-parameter Zipf model can be derived from the reconstructed equations of blood vessel network. By cumulating the bifurcation number of vessels at each level of equation (11) yields

$$N_m(L) = \sum_{i=1}^{m} N_i = N_1 \sum_{i=1}^{m} r_b^{m-1} = \frac{r_b^{m-1} - 1/r_b}{1 - 1/r_b}, \tag{14}$$

where $N_m(L)$ denotes the cumulative number of blood vessels from the first order ($m=1$) to the $m$th order ($m>1$), and $N_1=1$ is a coefficient indicating the number of blood vessels at the top level. From equation (12) it follows $r_l^{m-1}=L_1/L_m$, on both sides of which taking logarithms to the base $r_b$

yields

$$m-1 = \log_{r_b}(L_1/L_m)/\log_{r_b} r_l. \tag{15}$$

Substituting equation (15) into equation (14) yields

$$L_m = L_1(\frac{r_b}{r_b-1})^{\ln r_l/\ln r_b}[N_m(L) - \frac{1}{1-r_b}]^{-\ln r_l/\ln r_b}, \tag{16}$$

which are rearranged by transposition of terms and the base changing formula for logarithms are considered. Let $N_m(L)=r$ represent rank, and $L_m=L(r)$ represent the cumulative length of rank $r$. Thus equation (16) will be reduced to the following form

$$L(r) = C(r-\alpha)^{-d_z}, \tag{17}$$

in which $C=L_1[r_b/(r_b-1)]^{d_z}$, $\alpha=1/(1-r_b)$, $d_z=\ln r_l/\ln r_b$. Equation (17) is just the three-parameter Zipf model of the rank-size distributions, which differs in expression from the one- or two-parameter Zipf's law (Zipf, 1949). In fact, the thee-parameter Zipf model has been studied in various fields such as physics, geography, biology, and fractal theory (Chen and Zhou, 2003; Gabaix and Ibragimov, 2011; Gell-Mann, 1994; Mandelbrot, 1982; Winiwarter, 1983).

The three-parameter Zipf model suggests that the size distribution of vessel lengths comply with the generalized rank-size rule. The pure form of the rank-size rule indicates a scaling exponent equal to 1, while the generalized rank-size rule indicates a scaling exponent greater or less than 1. What is more, an allometric scaling relation between blood vessel length and radius can be derived from equations (12) and (13). Taking logs to the base $r_l$ on both sides of equation (13) gives

$$1-m = \log_{r_l}(R_m/R_1)/\log_{r_l} r_z. \tag{18}$$

Substituting equations (18) into equation (12) yields

$$\frac{L_m}{L_1} = (\frac{R_m}{R_1})^{\ln r_l/\ln r_v}, \tag{19}$$

which can be simplified in form as below

$$L_m = \mu R_m^b, \tag{20}$$

in which $\mu=L_1/R_1^b$ refers to a proportionality coefficient, and $b=\ln r_l/\ln r_z$ to allometric scaling exponent. So far, we have fulfilled the third task of reconstructing the models of blood vessel composition.

## 2.3 The fractal dimension of blood vessel networks

It can be proved that the three-parameter Zipf model and the allomety model are fractal models. These models can be employed to describe the hierarchy with cascade structure (Chen, 2012; Chen, 2014; Chen and Zhou, 2003). From equation (14) it follows

$$\frac{N_{m+1}(L)}{N_m(L)} = \frac{1-r_b^{m+1}}{1-r_b^m} = r_b + \frac{r_b-1}{r_b^m - 1}. \tag{21}$$

On the other hand, from equation (12) it follows

$$\frac{L_{m+1}}{L_m} = \frac{1}{r_l}. \tag{22}$$

In theory, a hierarchy of blood vessels comprise innumerable levels, and the number ratio $r_b>1$. In terms of equation (21) and (22), we have

$$D = -\lim_{m \to \infty} \frac{\ln[N_{m+1}(L)/N_m(L)]}{\ln[L_{m+1}/L_m]} = \frac{\ln r_b}{\ln r_l}, \tag{23}$$

which bears an analogy to the similarity dimension of fractals, and can be regarded as a fractal dimension of size distribution of blood vessel length. The parameter $d_z=1/D$ can be termed *Zipf dimension*, but it is actually the reciprocal of the fractal dimension of hierarchies of blood vessels.

Further, let $D_l$ represent the fractal dimension of blood vessel curves measured by $L_m$, and $D_z$ represent the fractal dimension of blood vessel diameter measured by $R_m$. Generally speaking, $D_z=1$. According to the geometric measure relation of fractal objects (Feder, 1988; Mandelbrot, 1982; Takayasu, 1990), we have an allometric relation

$$L_m \propto R_m^{D_l/D_z}. \tag{24}$$

Comparing equation (24) with equation (19) or equation (20) shows

$$b = \frac{D_l}{D_z} = \frac{\ln r_l}{\ln r_z}. \tag{25}$$

Obviously, if the scaling exponent *b* is not an integer or a ratio of two integers, it can be treated as fractal parameter associated with fractal dimension. In fact, the allometric scaling exponent implies the ratio of one fractal dimension to another fractal dimension. If $D_z=1$ as given, then $b=D_l$.

In this case, the allometric exponent is just the fractal dimension of blood vessels as fractal lines.

# 3 Empirical analysis

## 3.1 Data, methods, and results

An analogy between blood vessels and rivers has been drawn and fractal structure has been revealed from creatural hierarchies of arterial vessels (Jiang and He, 1989; Jiang and He, 1990). It was shown that Horton-Strahler's law can be used to describe human and dog's left and right auricular arteries as well as dog's epigastric rectus arteries (Liu *et al*, 1992). The basic measurements are blood vessel diameter, arterial length, and bifurcation number of different levels. Fitting equations (5), (9), and (10) to the biological data yield fractal parameter values (Table 1). Using the reconstructed model presented in the second section, we can evaluate more parameters such as fractal dimension, Zipf dimension, allometric scaling exponent, and the adjusted fractal dimension of the size distribution of arterial vessels (Table 2). In fact, given the radius of the aortic tube $R_1$ and the average cumulative length of arterial vessels $L_1$, all the parameters of the models proposed in this paper can be estimated, and we can build the three-parameter Zipf's model and the obtain the allometric scaling relation between vascular length and diameter.

**Table 1 The exponential models of the diameter (or radius), length, and bifurcation number of human and dog's arterial vessels at different levels**

| Species | Arterial tree | Original models | | | Reconstructed models | | |
|---|---|---|---|---|---|---|---|
| | | Diameter $d_m=d_0e^{\rho m}$ | Length $h_m=h_0e^{\sigma m}$ | Bifurcation $N_m=N_0e^{-\omega m}$ | Radius $R_m=R_1r_z^{m-1}$ | Length $L_m=R_1r_l^{m-1}$ | Bifurcation $N_m=N_1r_b^{1-m}$ |
| Human | Left auricular artery | $10.1601\times e^{0.4006m}$ | $0.1782\times e^{0.3638m}$ | $42442406.44 \times e^{-1.2142m}$ | $7.5831\times 1.4927^{m-1}$ | $0.2564\times 1.4388^{m-1}$ | $12603165.52 \times 3.3676^{1-m}$ |
| | Right auricular artery | $9.0657\times e^{0.4181m}$ | $0.0796\times e^{0.4711m}$ | $27809936.68 \times e^{-1.2427m}$ | $6.8857\times 1.5191^{m-1}$ | $0.1275\times 1.6018^{m-1}$ | $8026057.14 \times 3.4650^{1-m}$ |
| Dog | Left auricular artery | $8.8900\times e^{0.4119m}$ | $0.4712\times e^{0.2586m}$ | $5885725.45 \times e^{-1.1736m}$ | $6.6745\times 1.5097^{m-1}$ | $0.6103\times 1.2951^{m-1}$ | $1820170.18 \times 3.2336^{1-m}$ |
| | Right auricular artery | $8.9991\times e^{0.4236m}$ | $0.2807\times e^{0.3489m}$ | $889405.89 \times e^{-1.1949m}$ | $6.8728\times 1.5275^{m-1}$ | $0.3979\times 1.4175^{m-1}$ | $269307.46 \times 3.3026^{1-m}$ |

| | Epigastric rectus artery | $5.0746 \times e^{0.6360m}$ | $0.0571 \times e^{0.6083m}$ | $187327.00 \times e^{-1.3463m}$ | $4.7927 \times 1.8889^{m-1}$ | $0.1049 \times 1.8371^{m-1}$ | $48742.72 \times 3.8432^{1-m}$ |

**Note:** The original data come from Liu *et al* (1992). The models are reconstructed and the parameters are re-estimated in this paper. The variable $h$ denotes the branch length, and $L$ refers to the cumulative length.

The results show that Zipf's dimension $d_z$ values are fractional, and the allometric scaling exponent $b$ value are not the ratio of two integers (Table 2). This suggests that the systems of blood vessels possess the fractal structure similar to that of network of rivers, which can be modeled with the three-parameter Zipf's law and allometric scaling relation (Chen, 2009; Chen and Liu, 2001; Chen and Li, 2003). For both rivers and blood vessels, the Zipf's exponent is not close to 1. The special scaling exponent value of Zipf's distribution, 1, approximately appears in various rank-size distributions such as cities, firms, per capita incomes, and word frequencies (Bak, 1996; Carrol, 1982; Chen, 2008; Gabaix and Ioannides, 2004; Zhou, 1995).

**Table 2 The fractal dimension, Zipf's exponent, scaling factor, and other related parameters of human and dog's arterial trees**

| Type | Bifurcation ratio | Length ratio | Caliber ratio | Logarithmic ratio | | | Coefficient |
|---|---|---|---|---|---|---|---|
| | $r_b=N_m/N_{m+1}$ | $r_l=L_{m+1}/L_m$ | $r_z=R_{m+1}/R_m$ | $\omega=\ln r_b$ | $\sigma=\ln r_l$ | $\rho=\ln r_z$ | $C=L_1(r_b/(r_b-1))^{d_z}$ |
| Human left auricular artery | 3.3676 | 1.4388 | 1.4927 | 1.2142 | 0.3638 | 0.4006 | $1.1113 L_1$ |
| Human right auricular artery | 3.4650 | 1.6018 | 1.5191 | 1.2427 | 0.4711 | 0.4181 | $1.1378 L_1$ |
| Dog's left auricular artery | 3.2336 | 1.2951 | 1.5097 | 1.1736 | 0.2586 | 0.4119 | $1.0850 L_1$ |
| Dog's right auricular artery | 3.3026 | 1.4175 | 1.5275 | 1.1947 | 0.3489 | 0.4236 | $1.1111 L_1$ |
| Dog's epigastric rectus artery | 3.8432 | 1.8371 | 1.8889 | 1.3463 | 0.6083 | 0.6360 | $1.1459 L_1$ |

Continued Table 2

| Type | Zipf dimension | Fractal dimension | Scaling exponent | Inching parameter | Coefficient | Adjusted dimension |
|---|---|---|---|---|---|---|
| | $d_z=\ln r_l/\ln r_b$ | $D=\ln r_b/\ln r_l$ | $b=\ln r_l/\ln r_z$ | $\alpha=1/(1-r_b)$ | $\mu=L_1/R_1^b$ | $D^*=\ln r^*_b/\ln r^*_l$ |
| Human left auricular artery | 0.2996 | 3.3376 | 0.9081 | -0.4224 | $L_1/R_1^{0.91}$ | 2.8958 |
| Human right auricular artery | 0.3791 | 2.6379 | 1.1268 | -0.4057 | $L_1/R_1^{1.13}$ | 2.3071 |
| Dog's left auricular artery | 0.2204 | 4.5383 | 0.6278 | -0.4477 | $L_1/R_1^{0.63}$ | 3.8887 |
| Dog's right auricular artery | 0.2920 | 3.4242 | 0.8237 | -0.4343 | $L_1/R_1^{0.82}$ | 2.9547 |
| Dog's epigastric rectus artery | 0.4518 | 2.2132 | 0.9565 | -0.3517 | $L_1/R_1^{0.96}$ | 1.9841 |

**Note:** The parameters are estimated by the original data coming from Liu *et al* (1992).

## 3.2 Analysis and findings

The theoretical derivations and empirical analysis show that the fractal structure of a hierarchy of blood vessels is similar to that of a network of rivers. The mathematical models of river networks can be applied to the hierarchy of blood vessels. However, where the blood vessels are concerned, how to understand the fractal dimension of the rank-size distribution is still a problem. As indicated above, the fractal dimension of hierarchies of blood vessels is $D=1/d_z$. In fact, its meaning is more than this. From equations (11) and (12) it follows

$$\frac{N_m}{N_1} = \left(\frac{L_m}{L_1}\right)^{-\ln r_b / \ln r_l}. \qquad (26)$$

Suppose that the fractal dimension of the measurement $N_m$ is $D_b$. According to the fractal measure relation (Takayasu, 1990), we have

$$N_m \propto L_m^{-D_b/D_l}. \qquad (27)$$

Comparing equation (26) with equation (27) shows

$$D = \frac{1}{d_z} = \frac{D_b}{D_l}. \qquad (28)$$

This suggests that the fractal dimension of the rank-size distribution of blood vessels (fractal hierarchy) $D$ is the ratio of the fractal dimension of the spatial structure of blood vessels (fractal network) $D_b$ to the average dimension of the blood vessels (fractal lines) $D_l$.

The parameter relation is revealing for us to understand the fractal dimension of the systems of blood vessels. Since $D_b \leq 3$, $D_l \geq 1$, we should have $D=1/d_z \leq D_b \leq 3$, and thus $d_z \geq 1/3$. On the other, in reality, $D_b \geq 2$, $D_l \approx 1$, thus $D=1/d_z \geq 2$. This suggests that the $D$ value ranges from 2 to 3 in principle, and the upper limit ($D=3$) is stricter than the lower limit ($D=2$). However, Table 2 shows that some calculated values of the fractal dimension are greater than 3, namely, $D>3$. The reasons may be as follows. First, the method of determining the bifurcation ratio is by analogy with the approach proposed by Strahler (1952, 1958). This method is not based on scaling invariance completely and may overestimate the $r_b$ value, and therefore the fractal dimension by the formula $D=\ln r_b/\ln r_l$ is greater than its actual value. Second, the dimension of hierarchies ($D=1/d_z$) is a type of similarity dimension that differs from the box dimension. The box dimension comes between the topological dimension and the Euclidean dimension of the embedding space of a fractal. However, the similarity dimension may exceed the Euclidean dimension of the embedding space. In a sense, the similarity dimension is defined for the generalized space rather than the real space (Chen, 2008; Chen, 2014). In short, the calculations are understandable and thus can be acceptable.

The fractal dimension values shown above can be adjusted by using scaling idea. According to Strahler's method of bifurcation structure, the bifurcation ratio is not equal to $r_b=N_{m-1}/N_m$, but it is as below

$$r_b = \frac{N_m+1}{N_{m+1}} = r_b^* + \frac{1}{N_{m+1}}, \tag{29}$$

where $b=N_m/N_{m+1}$ denotes the bifurcation ratio based on the bottom-up hierarchy. Actually, Strahler's bifurcation ratio $r_b$ is not strictly based on linear scaling, and $r_b>r_b^*$. The scaling ratio is $r_b^*$ instead of $r_b$. Generally speaking, $N_1$ is very large. If $N_m=N_2$, then we have $r_b=r_b-1/N_2=r_b(1-1/N_1)\approx r_b$; if $N_m=N_M=1$ as given, then $r_b=r_b-1/N_M=r_b-1$. The average value of the two extreme results is

$$r_b^* = \frac{2r_b-1}{2} = r_b - \frac{1}{2}. \tag{30}$$

Based on equation (30), the fractal dimension can be adjusted by the formula $D^*=\ln r_b^*/\ln r_l$.

Using the formulae given above, we can re-determine the bifurcation ratio and then adjust the fractal dimension. The results are displayed in the last column of Table 2. An interesting discovery is that the similarity dimension values of human arteries come between 2 and 3, but those of dog's arteries may go beyond the upper limit 3 or the lower limit 2. This accounts for the differences between human beings and dogs. There exists analogy between the physical structure of human beings and that of dogs. However, compared with dog's bodies, human bodies more conform to natural rules. The fractal dimension values of human bodies fall into the reasonable range, that is, 2<*D*<3. In contrast, the fractal dimension values of dogs' bodies are sometimes greater than 3 and sometimes less than 2, namely, *D*<2 or *D*>3 for time to time.

## 4 Questions and discussion

The ramiform patterns of blood vessels are similar to the branchlike patterns of rivers. The similarities and differences between the patterns of blood vessels and those of rivers are helpful for our understanding the self-organizing processes of natural and human systems. By the mathematical transform and empirical analysis, the basic properties of hierarchies of blood vessels can be revealed as follows. The first is *recurrence*. The recursion is shown by equations (5), (9), and (10). This suggests a self-similarity or self-affinity of blood vessel organization. The second is *symmetry*. The recursive structure suggests the translational symmetry of scale and scaling symmetry of allometric growth. The third is *imitation*. The organisms such as human bodies seem to give an imitation of its geographical environment in the course of evolution. Due to this imitation, there emerge similarities between physical structure and physiographic configuration.

However, the correspondence between the models of blood vessels and the laws of river composition is not complete. By analogy, both the river laws and blood vessel models can be developed. First, the linear scaling law of the blood vessel diameter has no counterpoint of rivers. We need a model to describe river cross sections. Second, the linear scaling law of river drainage area bears no counterpoint of blood vessels. We need a model to characterize the catchment volumes of blood vessels. Thus we have a one-to-one correspondence between blood vessel measurements and rivers measurements that are tabulated as follows (Table 3). Note that the cross section area of blood vessels can be replaced by diameter or radius. Similarly, the cross section areas of rivers can be substituted with the depths at thalweg points.

Table 3 The one-to-one correspondence between blood vessel measurements, rivers measurements, and city measurements

| Measurement | Blood vessels | Rivers | Cities |
| --- | --- | --- | --- |
| Number | Bifurcation number | Bifurcation number | City number of similar size |
| Size | Length | Length | Population |
| **Shape 1** | Catchment volume | Drainage area | Service area |
| **Shape 2** | Cross section | Cross section | Urbanized area |

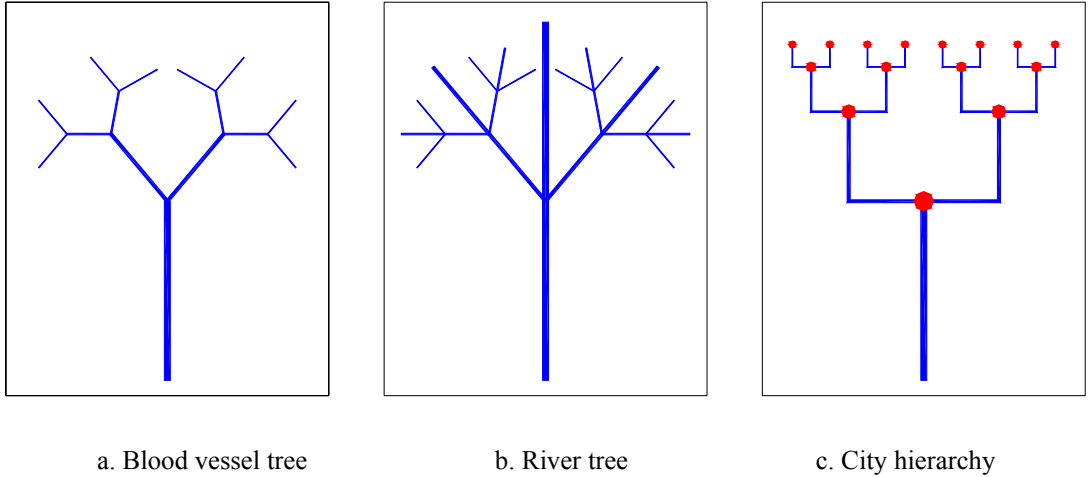

a. Blood vessel tree     b. River tree     c. City hierarchy

**Figure 1 Sketch maps of models of bottom-up hierarchies of blood vessels, rivers, and cities (the first four levels)**

    The analogy between the hierarchies of blood vessels and networks of rivers can be generalized to the analogy between the hierarchies of blood vessels and hierarchies of cities because urban evolution is associated with river patterns (Figure 1). The geographical locations of urban growth bear three basic properties (Chen, 1995). The first is waterfront effect. City development depends on rivers or lakes, especially rivers. The second is edge effect. A city is often located on the border between two or more different regional systems. The third is counterpart effect. Landform is fractal (Rodriguez-Iturbe and Rinaldo, 2001; Turcotte, 1997). However, the fractal form of some parts degenerate because of weathering and fluviation. Cities are self-similar or self-affine (Batty and Longley, 1994; Frankhauser, 1994). A fractal city usually develops in the fractal-free area within a fractal terrain. Rivers influence the interaction, spatial structure, rank-size distribution, and differentiation of functions of cities. The nonlinear correlation between rivers and cities results in similarity between hierarchies of cities and network of rivers. The analogy between rivers and urban places has been studied for a long time, and the similarities between rivers and cities have been revealed at several aspects (Chen, 2009; Chen and Zhou, 2006; Krugman, 1996; Woldenberg, 1968; Woldenberg and Berry, 1967). By analogy, the exponential laws of blood

vessels and rivers can be applied to the number, size, and shape of cities (Table 4). From these exponential functions it follows a set of power laws indicative of fractals, rank-size distributions, and allometric scaling. Thus we can gain an insight into the hierarchies of blood vessels in the right perspective.

Table 4 The linear scaling laws of number, size, and shapes of blood vessels, rivers, and cities

| Measurement | Blood vessels | Rivers | Cities |
| --- | --- | --- | --- |
| Number (bifurcation, quantity) | $N_m = N_1 r_b^{m-1}$ | $N_m = N_1 r_b^{m-1}$ | $N_m = N_1 r_n^{m-1}$ |
| Size (length, population) | $L_m = L_1 r_l^{1-m}$ | $L_m = L_1 r_l^{1-m}$ | $P_m = P_1 r_p^{1-m}$ |
| Shape 1 (volume, area) | $V_m = V_1 r_v^{1-m}$ | $A_m = A_1 r_a^{1-m}$ | $A_m = A_1 r_a^{1-m}$ |
| Shape 2 (section, area) | $S_m = S_1 r_s^{1-m}$ | $S_m = S_1 r_s^{1-m}$ | $U_m = U_1 r_u^{1-m}$ |

**Note:** For blood vessels, $N_m$ refers to the bifurcation number of vessels of order $m$, $L_m$ to the average length of vessels of order $m$, $V_m$ to the average catchment volume of vessels of order $m$, and $S_m$ to the average cross-sectional area of vessels of order $m$; For rivers, $N_m$ refers to the bifurcation number of rivers of order $m$, $L_m$ to the average length of rivers of order $m$, $A_m$ to the average drainage area of rivers of order $m$, and $S_m$ to the average cross-sectional area of rivers of order $m$; For cities, $N_m$ refers to the number of cities of order $m$, $P_m$ to the average population of cities of order $m$, $A_m$ to the average service area of cities of order $m$, and $U_m$ to the average urbanized area of cities of order $m$.

In this work, the hierarchy of blood vessels is treated as a eudipleural structure by means of statistical average. In other words, the blood vessel hierarchy is modeled by monofractal geometry indicative of self-similarity. The pattern is of bilateral symmetry. However, the real hierarchy of blood vessels is asymmetric, and it should be modeled with self-affine fractal geometry or multifractal geometry (Mainzer, 1994). There are significant similarities and differences between self-similar fractals (monofractal), self-affine fractals (sometimes takes on bi-fractals), and multifractals (Table 5). A self-similar fractal object is isotropic, and it shows the uniform growth probability of different fractal units and the same growth rate in different directions (Figure 2(a)). A self-affine fractal object is anisotropic, and it displays the uniform growth probability of different fractal units but the different growth rate in different directions (Figure 2(b)). In a 2-dimensional space, a self-affine fractal pattern bears two fractal dimension values. A multifractal object displays different growth probabilities of different fractal units (Figure 2(c)). In a 2-dimensional space, a multifractal pattern bears numberless fractal dimension value. Where average values of different levels are concerned, the hierarchy of blood vessels can be described with the three-parameter Zipf law, which indicates monofractal patterns. In fact, the three-parameter Zipf model can be generalized to multifractal measures (Chen and Zhou, 2004). The multifractal Zipf model can be applied to the asymmetric hierarchy of blood vessels in empirical analyses. What is more, the Zipf dimension of self-similar hierarchy is in fact the ratio of two fractal dimensions rather than the reciprocal of a fractal dimension (Chen, 2014; Chen and Zhou, 2006). As space is limited, the related issues remains to be clarified in future studies.

**Table 5 The similarities and differences between self-similar fractals, self-affine fractals, and multifractals**

| Item | Self-similar fractals | Self-affine fractals | Multifractals |
|---|---|---|---|
| **Spatial spread** | Isotropy | Anisotropy | Isotropy or anisotropy |
| **Growth probability** | Uniform | Uniform | Nonuniform |
| **Pattern** | Symmetry | Asymmetry | Symmetry or asymmetry |
| **Fractals** | Monofractals | Bi-fractals | Multiscaling fractals |

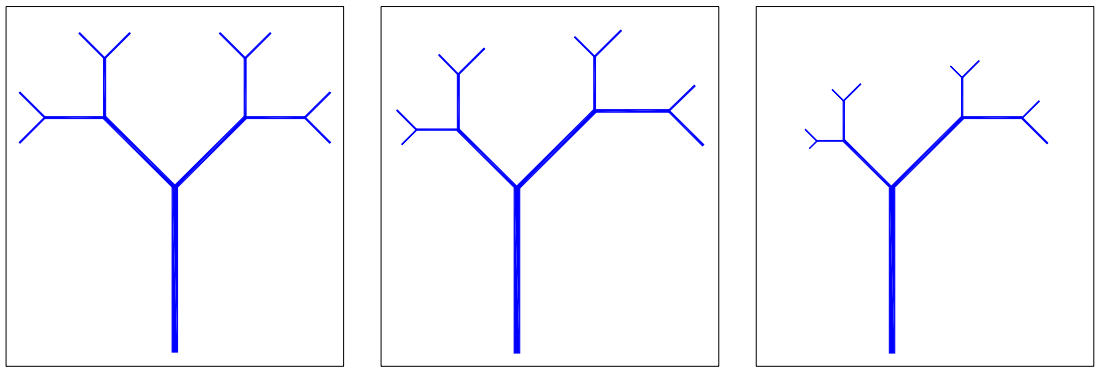

a. Self-similar structure      b. Self-affine structure      c. Multifractal structure

**Figure 2 Sketch maps of three fractal models of hierarchies of blood vessels (the first four levels)**
**Note:** For the self-similar fractal structure, the scale ratio is $r_f=1/2$; for the self-affine fractal structure, the scale ratios are $r_{long}=2/3$ and $r_{short}=1/2$; for the multifractal structure, the scale ratios are $r_{long}=3/5$ and $r_{short}=1- r_{long}=2/5$.

# 5 Conclusions

The development of organisms depends on geographical environments, and both organisms and its geographical environments follow the same natural laws such as spatial, allometric, and hierarchical scaling laws. Studies on analogy between creatures and geographical environments are revealing for us to understand the deep structure and complex dynamics of natural and social systems. The novelty of this paper is as below: first, the models of hierarchies of blood vessels are developed by mathematical reasoning; second, the estimation formula of the similarity dimension based on Strahler's classification is improved by scaling analysis; third, the linear scaling laws of hierarchies of blood vessels are generalized by analogy with rivers and cities; fourth, an interesting discovery is made on the similarities and differences between human beings and dogs.

The main points of this study can be summarized as follows. **First, there exist analogy of cascade structure between the self-similar hierarchy of blood vessels and the fractal network of rivers.** The analogy can be generalized to the similarity between urban systems and blood vessel hierarchies. The blood vessel length corresponds to the river length and city population size, the drainage volume of a blood vessel corresponding to the catchment area of a river and the service area of a city, and the bifurcation number of blood vessels corresponds to the bifurcation number and rivers and the city number of certain class. All these hierarchical structures can be described with a set of exponential functions indicative of linear scaling laws, from which it follows a series of power laws indicating fractals or allometry. **Second, the deep structure of a hierarchy of blood vessels can be characterized by the three-parameter Zipf's law and allometric scaling laws.** As indicated above, the blood vessel hierarchies can be described with a set of exponential laws. The exponential laws represent linear scaling laws, from which we can derive a set of nonlinear scaling laws such as the three-parameter Zipf's law and allometric scaling laws. The three-parameter Zipf's law reflects the fractal property of the rank-size distributions of blood vessel lengths, and the allometric scaling laws reflects the fractal dimension relationships between blood vessel length and vessel diameter as well as the drainage volume of a blood vessel. The dynamics behind the deep structure remains to be researched in future. **Third, the analogy studies between the fractal structure of human beings and that of other mammals can reveal the similarities and differences between human and animals.** The similarity dimension values of human arteries fall into the rational range, coming between 2 and 3. However, the dimension values of dog's arteries sometimes go beyond the lower limit 1 or the upper limit 3. This suggests that human bodies more conform to natural rules than dogs' bodies. The conclusion can be generalized to other fractal phenomena. An inference is that the fractal dimension values of a river network or urban system with sound structure will fall into a reasonable range ($1<D<3$); on the contrary, the fractional dimension values of an river network or urban system with unsound structure will go beyond the reasonable range ($D<1$ or $D>3$).

# Acknowledgement

This research was supported financially by the National Natural Science Foundation of China (Grant no. 41171129). The support is gratefully acknowledged.

Horton RE (1945). Erosional development of streams and the drainage basins: Hydrophysical approach to quantitative morphology. *Bulletin of the Geological Society of America*, 56(3): 275-370

Jiang B, Yao XB (2010 eds.). *Geospatial Analysis and Modeling of Urban Structure and Dynamics*. New York: Springer

Jiang ZL, He GC (1989). Geometrical morphology of the human coronary arteries. *Journal of Third Military Medical University*, 11(2): 85-91 [In Chinese]

Jiang ZL, He GC (1990). Geometrical morphology of coronary arteries in dog. *Chinese Journal of Anatomy*, 13(3): 236-241 [In Chinese]

Krugman P (1996). Confronting the mystery of urban hierarchy. *Journal of the Japanese and International economies*, 10(4): 399-418

LaBarbera P, Rosso R (1989). On the fractal dimension of stream networks. *Water Resources Research*, 25(4): 735-741

Lee TD (1999). *Arts and Science* (*Yi Shu Yu Ke Xue*). Hangzhou: Zhejiang Literature and Art Publishing House [ Edited by H.Z. Liu in Chinese]

Li MH, Ying DJ, Zhang JS (1992). Studies on fractals in biomedicine (*Sheng wu yix ue zhong de fen xing yan jiu*). *Nature Magazine*, 15(8): 59 2-596 [In Chinese]

MacDonald N (1983). *Trees and Networks in Biological Models*. Chichester, UK: John Wiley and Sons

Mainzer K (1994). *Thinking in Complexity: The Complex Dynamics of Matter, Mind, and Mankind*. Berlin: Springer-Verlag

Mandelbrot BB (1982). *The Fractal Geometry of Nature.* New York: W. H. Freeman and Company

Nelson TR, Manchester DK (1988). Modeling of lung morphogenesis using fractal geometry. *IEEE Transactions on Medical Imaging*, 7: 321-327

Pumain D (2006 ed). *Hierarchy in Natural and Social Sciences*. Dordrecht: Springer

Rodriguez-Iturbe I, Rinaldo A (2001). *Fractal River Basins: Chance and Self-Organization*. New York: Cambridge University Press

Schumm SA (1956). Evolution of drainage systems and slopes in badlands at Perth Amboy, New Jersey. *Geological Society of America Bulletin*, 67(5), 597-646

Strahler AE (1952). Hypsometric (area-altitude) analysis of erosional topography. *Geological Society of American Bulletin*, 63(11): 1117-1142

Strahler AN (1958). Dimensional analysis applied to fluvially eroded landform. *Geological Society of*20